\documentclass[aps,pra,twocolumn,superscriptaddress,nofootinbib]{revtex4-1}
\usepackage{amsmath}
\usepackage{amssymb}
\usepackage{graphicx}
\usepackage{mathrsfs}
\usepackage{ntheorem}
\usepackage{mathtools}
\usepackage{enumerate}
\usepackage{enumitem}
\usepackage{times,txfonts}
\usepackage{subfigure}
\usepackage{upgreek}
\usepackage{tikz}
\usepackage{soulutf8}
\usepackage{xcolor}
\usepackage[colorlinks,bookmarksopen,bookmarksnumbered,citecolor=blue, linkcolor=blue, urlcolor=blue]{hyperref}
\newcommand{\ket}[1]{|#1\rangle}
\newcommand{\bra}[1]{\langle #1|}
\newcommand{\Tr}{\mathrm{Tr}}

\newcommand{\abs}[1]{\lvert #1\rvert}

\def\CC{{\rm\kern.24em \vrule width.04em height1.46ex depth-.07ex \kern-.30em C}}
\def\RR{{\rm\kern.24em \vrule width.04em height1.46ex depth-.07ex
\kern-.30em R}}
\def\P{{\rm I\kern-.25em P}}
\colorlet{hlcolor}{yellow!40}

\begin{document}

\title{Incoherent Operations Enable State Transformations Impossible under Dephasing-covariant Incoherent Operations}

\author{C. L. Liu}
\email{clliu@sdu.edu.cn}
\affiliation{School of Information Science and Engineering, Shandong University, Qingdao 266237, China}
\date{\today}

\begin{abstract}
We show that incoherent operations (IOs) can achieve the state transformations that are forbidden under dephasing-covariant incoherent operations (DIOs), thereby resolving the open problem posed by Chitambar and Gour [Phys. Rev. Lett. 117, 030401 (2016)]. We further demonstrate that no set of IO monotones suffices to characterize state convertibility under strictly incoherent operations (SIOs), and that monotones common to IOs and DIOs are insufficient to characterize convertibility under DIOs.
\end{abstract}
\maketitle

\section{Introduction}
Quantum coherence is a fundamental feature of quantum mechanics that distinguishes the quantum world from the classical one. It plays a crucial role in quantum information processing and is central to various fields \cite{Nielsen}, including quantum computation \cite{Shor,Grover}, quantum cryptography \cite{Bennett}, quantum metrology \cite{Giovannetti}, and quantum biology \cite{Lambert}. The resource theory of coherence has attracted significant attention due to the rapid development of quantum information science \cite{Aberg2006AX,Baumgratz2014PRL,Streltsov2017RMP,Hu2018PP}. This theory not only provides a rigorous framework for quantifying coherence but also offers a new perspective for understanding its role in quantum systems.

Every quantum resource theory has two fundamental elements: free states and free operations \cite{Chitambar2019RMP}. In the resource theory of coherence, the free states are those that are diagonal in a fixed reference basis. The definition of free operations is not unique, and several classes have been introduced based on different operational considerations. These include maximally incoherent operations (MIOs) \cite{Aberg2006AX}, which never generate coherence from incoherent states; dephasing-covariant incoherent operations (DIOs) \cite{Chitambar2016PRA,Chitambar2016PRL,Marvian2016PRA}, which form the largest class of operations that cannot detect coherence in any input state; incoherent operations (IOs) \cite{Baumgratz2014PRL}, which forbid the creation of coherence from incoherent states even probabilistically; and strictly incoherent operations (SIOs) \cite{Winter2016PRL,Yadin2016PRX}, which neither create nor exploit coherence.

A fundamental task in the resource theory of coherence concerns the differences among these classes of free operations, particularly regarding the state transformations they allow \cite{Hu2018PP,Streltsov2017RMP,Chitambar2016PRA,Chitambar2016PRL}. This comparison was first carried out in Refs. \cite{Chitambar2016PRA,Chitambar2016PRL} and has since received considerable attention \cite{Liu2020PRA,Liu2022PRA,Streltsov2017PRL,Fang2018PRL,Lami2019PRL,Liu2019PRL,Regula2018PRL,Zhao2019IEEE,Lami2020IEEE,Bu2017QIC}. From the inclusion structure of these classes, one has \(\mathrm{SIOs} \subset \mathrm{IOs} \subset \mathrm{MIOs}\) and \(\mathrm{SIOs} \subset \mathrm{DIOs} \subset \mathrm{MIOs}\), while IOs and DIOs are known to be incomparable. In terms of state convertibility, MIOs form the strongest class, as they can increase the coherence rank of certain states, which is impossible for other free operations \cite{Chitambar2016PRA}. It was further shown in Refs. \cite{Liu2020PRA,Liu2019PRL} that both IOs and DIOs strictly outperform SIOs in state-conversion power. Since IOs and DIOs are not ordered by set inclusion, there is no reason to expect one class to dominate the other operationally. Regarding advancements in this issue, it is shown that DIOs and IOs possess equivalent power for certain tasks, such as transforming qubit states  \cite{Chitambar2016PRA} and pure coherent states \cite{Regula2020PRR}. Then, Ref. \cite{Bu2017QIC} demonstrated that DIOs can be strictly more powerful than IOs for certain state transformations. This naturally raises the converse question: whether IOs can, in turn, outperform DIOs for some state transformations. This problem was first raised in Refs. \cite{Chitambar2016PRA,Chitambar2016PRL} and has remained open until now.

Here, we address this problem by explicitly constructing a state transformation achievable with IOs but impossible under both SIOs and DIOs. This proves that IOs can be strictly stronger than DIOs, completing the operational comparison between these two incomparable classes. Building on this construction, we further show that IO-monotones alone are insufficient to characterize SIO convertibility, and that even the intersection of IO and DIO monotones fails to characterize DIO convertibility. These results reveal a necessary distinction between the sets of monotones for different operational classes and clarify the limitations of using monotones inherited from larger sets of operations.

\section{Resource theory of coherence}

To present our results clearly, we recall the basic framework of the resource theory of coherence \cite{Baumgratz2014PRL,Streltsov2017RMP}.

Let $\mathcal{H}$ be the Hilbert space of a $d$-dimensional quantum system, and let \(\{\ket{j}\}_{j=0}^{d-1}\) denote a fixed reference basis, chosen according to the physical context. Coherence of a quantum state is always measured with respect to this basis. A general density operator can be written as \(\rho=\sum_{ij}\rho_{ij}\ket{i}\bra{j}\), where \(\rho_{ij}\) are the matrix elements. A state is called incoherent if its density operator is diagonal in the reference basis. The set of all incoherent states is denoted by \(\mathcal{I}\) and hence any \(\delta\in\mathcal{I}\) takes the form \(\delta=\sum_i\delta_i\ket{i}\bra{i}\). States that are not diagonal in the basis are referred to as coherent states. A pure state is written as \(\ket{\varphi}=\sum_{j=0}^{d-1} c_j\ket{j}\) with coefficients \(c_j\), and its density operator is \(\varphi=\ket{\varphi}\bra{\varphi}\).

We now introduce the three main classes of free operations used in this work. An incoherent operation \cite{Baumgratz2014PRL} is a completely positive trace-preserving (CPTP) map $\Lambda(\rho)=\sum_\mu K_\mu\rho K_\mu^\dagger$ whose Kraus operators satisfy \(\sum_\mu K_\mu^\dagger K_\mu=I\) and
\begin{eqnarray}
K_\mu\mathcal{I}K_\mu^\dagger\subset\mathcal{I}
\end{eqnarray}
for every \(\mu\). The latter condition guarantees that each Kraus operator maps an incoherent state to another incoherent state. One can show that an incoherent Kraus operator has at most one nonzero entry in each column. A dephasing-covariant incoherent operation \cite{Chitambar2016PRL, Chitambar2016PRA, Marvian2016PRA} is a CPTP map that commutes with the dephasing operation \(\Delta\), defined by \(\Delta(\rho)=\sum_{j=0}^{d-1}\bra{j}\rho\ket{j}\ket{j}\bra{j}\). That is, a map \(\Lambda(\cdot)\) is a DIO if and only if
\begin{eqnarray}
[\Lambda,\Delta]=0.
\end{eqnarray}
This commutation condition ensures that the operation does not detect the coherence present in the input state. An strictly incoherent operation \cite{Yadin2016PRX, Winter2016PRL} is a CPTP map whose Kraus operators satisfy both
\begin{eqnarray}
K_\mu\mathcal{I}K_\mu^\dagger\subset\mathcal{I}~\text{and}~K_\mu^\dagger\mathcal{I}K_\mu\subset\mathcal{I}
\end{eqnarray}
for all \(\mu\). In other words, each $K_\mu$ and its adjoint must map incoherent states to incoherent states.

A functional \( C(\rho) \) qualifies as a coherence measure within the resource theories defined by IOs, DIOs, or SIOs provided it satisfies the following four requirements \cite{Baumgratz2014PRL}. The conditions are stated here using IOs as the representative class of free operations. (I) Non-negativity. $C(\rho)\ge 0$ for all states \(\rho\), and \(C(\delta)=0\) if and only if \(\delta\in\mathcal{I}\). (II) Monotonicity under free operations. For any IO \(\Lambda\), \(C(\rho)\ge C(\Lambda(\rho))\).
(III) Monotonicity under selective measurements.  If \(\Lambda(\rho)=\sum_\mu K_\mu\rho K_\mu^\dagger\) is an IO with outcomes labeled by \(\mu\), then \(C(\rho)\ge\sum_\mu p_\mu C(\rho_\mu)\), where \(p_\mu=\Tr(K_\mu\rho K_\mu^\dagger)\) and \(\rho_\mu=K_\mu\rho K_\mu^\dagger/p_\mu\). (IV) Convexity. For any ensemble \(\{\rho_n\}\) and any probability distribution \(\{q_n\}\), there is $\sum_n q_n C(\rho_n)\ge C\left(\sum_n q_n\rho_n\right)$. If a functional satisfies the first two conditions (I) and (II), we refer to it as a coherence monotone.

\section{Increasing $C_m(\rho)$ by using IOs}

With the above notions in place, we now present the central findings of this work, summarized in the following theorem.

\emph{Theorem} 1.--There exist quantum state transformations achievable by IOs that cannot be realized by SIOs or DIOs.

\emph{Proof.}--Let us consider the coherence measure
\begin{eqnarray}
C_m(\rho)=\frac{1}{d}\uplambda_{\max}\left((\Delta\rho)^{-1/2}\rho(\Delta\rho)^{-1/2}\right)-\frac{1}{d},
\label{maximum}
\end{eqnarray}
where, for a given state \(\rho = \sum_{ij} \rho_{ij}\ket{i}\bra{j}\), \(\Delta\rho = \sum_i \rho_{ii}\ket{i}\bra{i}\) denotes its diagonal part, and \((\Delta\rho)^{-1/2}\) is the inverse square root of \(\Delta\rho\), defined element-wise as \((\Delta\rho)^{-1/2}_{ii} = \rho_{ii}^{-1/2}\) if \(\rho_{ii}\neq 0\) and \(0\) otherwise. Here, \(\uplambda_{\rm max}(A)\) denotes the largest eigenvalue of a matrix \(A\) \cite{Liu2023PRA}. It has been established that \(C_m\) is a coherence monotone under DIOs \cite{Chitambar2016PRA} and a valid coherence measure under SIOs \cite{Liu2023PRA}. Consequently, for any initial state \(\rho_{\rm in}\), the action of either a DIO or an SIO cannot increase \(C_m\); that is, the output state \(\rho_{\rm out} := \Lambda(\rho_{\rm in})\) always satisfies
\begin{eqnarray}
C_m(\rho_{\rm in}) \ge C_m(\rho_{\rm out}).
\end{eqnarray}

Let us examine whether IOs can increase $C_m(\rho)$. While monotonicity is guaranteed under SIO and DIO, the behavior of $C_m(\rho)$ under general IO is not a priori obvious and must be checked explicitly. In the $2$-dimensional case, IOs, SIOs, and DIOs are known to be equivalent for state transformations \cite{Chitambar2016PRA}. Consequently, no IO can increase $C_m(\rho)$ in the $2$-dimensional case. Therefore, to observe a potential increase, we must examine at least the $3$-dimensional case.

To understand how such an increase can occur, it is instructive to rewrite $C_m(\rho)$ in the form of a Rayleigh quotient \cite{Horn}, i.e., for $\ket{\varphi}$ not being a null vector,
\begin{eqnarray}
C_m(\rho)=\frac{1}{d}\left(\max_{\ket{\varphi}}
\frac{\bra{\varphi}\rho\ket{\varphi}}{\bra{\varphi}\Delta\rho\ket{\varphi}}-1\right).
\label{Rayleigh}
\end{eqnarray}
By writing $\rho = \Delta\rho + \chi$, where $\chi$ contains only the off-diagonal elements of $\rho$, we obtain
\begin{eqnarray}
\frac{\bra{\varphi}\rho\ket{\varphi}}{\bra{\varphi}\Delta\rho\ket{\varphi}}
=1+\frac{\bra{\varphi}\chi\ket{\varphi}}{\bra{\varphi}\Delta\rho\ket{\varphi}}.
\label{off-diagonal}
\end{eqnarray}
This means $C_m(\rho)$ quantifies the maximal relative enhancement provided by $\chi$ with respect to the classical population distribution $(\rho_{11},\cdots,\rho_{dd})$. With these considerations, for a Kraus representation $\{K_\mu\}$ of an incoherent operation $\Lambda(\cdot)$ and an input state $\rho_{\mathrm{in}}$, the output $\rho_{\mathrm{out}}=\sum_\mu K_\mu\rho_{\mathrm{in}}K_\mu^{\dagger}$ satisfy $C_m(\rho_{\mathrm{out}})>C_m(\rho_{\mathrm{in}})$ if there exists a nonzero vector $\ket{\varphi}$ such that
\begin{eqnarray}
\frac{\bra{\varphi}\rho_{\text{out}}\ket{\varphi}}{\bra{\varphi}\Delta\rho_{\text{out}}\ket{\varphi}}> \max_{\ket{\psi}}\frac{\bra{\psi}\rho_{\text{in}}\ket{\psi}}{\bra{\psi}\Delta\rho_{\text{in}}\ket{\psi}}.
\label{increase}
\end{eqnarray}
The above analysis indicates that Eq. (\ref{increase}) can be realized through two distinct mechanisms:
(i) constructive interference of several off-diagonal elements of $\rho_{\text{in}}$ in the output, which increases $\bra{\psi}\chi_{\text{in}}\ket{\psi}$ for the optimal $\ket{\psi}$;
(ii) selective reduction of populations $(\rho_{11},\cdots,\rho_{dd})$ precisely for indices where the optimal $\ket{\psi}=\sum_jc_j\ket{j}$ exhibits large $\abs{c_j}$, thereby decreasing the denominator in Eq. (\ref{increase}).

To explicitly realize Eq.~(\ref{increase}), we now convert the above qualitative mechanisms into a concrete construction.
Guided by the two effects identified below Eq.~(\ref{increase}), our objective is to engineer an input state and an incoherent operation that jointly enhance the Rayleigh quotient in Eq.~(\ref{Rayleigh}). We proceed in three steps: (i) We choose an input state whose coherence is mainly concentrated in a $2$-dimensional subspace, so that the optimal Rayleigh quotient is primarily determined by interference within this subspace. (ii) We design incoherent Kraus operators that coherently mix the basis states in this subspace, thereby generating constructive interference among the relevant off-diagonal terms in the output state and increasing the numerator of Eq.~(\ref{increase}). (iii) We exploit the population redistribution induced by the Kraus operators to selectively reduce the diagonal weights associated with the optimal support, which decreases the denominator of Eq.~(\ref{increase}). Through this combined effect-enhancement of the numerator via interference and suppression of the denominator via population redistribution the quantity $C_m(\rho)$ can increase under an incoherent operation.

Implementing the above strategy, we arrive at the following explicit example: Consider the input state
\begin{eqnarray}
\rho_{\mathrm{in}} =
\begin{pmatrix}
\frac12 + \frac{\sqrt2}{20} &
\frac1{4\sqrt2} - \frac{i}{20} &
-\frac1{4\sqrt2} + \frac{i}{80} \\[4pt]
\frac1{4\sqrt2} + \frac{i}{20} &
\frac14 - \frac{\sqrt2}{40} &
\frac1{8\sqrt2} + \frac{i}{100} \\[4pt]
-\frac1{4\sqrt2} - \frac{i}{80} &
\frac1{8\sqrt2} - \frac{i}{100} &
\frac14 - \frac{\sqrt2}{40}
\end{pmatrix}.
\label{input_state}
\end{eqnarray}
A direct numerical evaluation of the coherence measure $C_m(\rho)$ for $\rho_{\mathrm{in}}$ yields
\begin{eqnarray}
C_m(\rho_{\mathrm{in}}) \approx 0.1853.
\end{eqnarray}
An explicit IO of the form \(\Lambda(\cdot) = K_1 \cdot K_1^{\dagger} + K_2 \cdot K_2^{\dagger}\) is constructed with
\begin{eqnarray}
K_{1}=
\begin{pmatrix}
0 & \tfrac1{\sqrt2} & \tfrac1{\sqrt2} \\[2pt]
0 & 0 & 0 \\[2pt]
0 & 0 & 0
\end{pmatrix},
\qquad
K_{2}=
\begin{pmatrix}
1 & 0 & 0 \\[2pt]
0 & \tfrac1{\sqrt2} & -\tfrac1{\sqrt2} \\[2pt]
0 & 0 & 0
\end{pmatrix},
\label{incoherent_operation}
\end{eqnarray}
which simultaneously exploits both mechanisms outlined below Eq. (\ref{increase}). It is straightforward to verify that $K_1^{\dagger}K_1 + K_2^{\dagger}K_2 = I$ and that $\Lambda(\cdot)$ is an IO, since for any incoherent state $\delta$, each term \(K_j\delta K_j^{\dagger}\) remains incoherent. However, \( \Lambda(\cdot) \) is neither an SIO nor a DIO.
The former follows from \( K_1^\dag|1\rangle\langle 1|K_1 \notin \mathcal{I} \), while the latter is demonstrated by the fact that $\Lambda\circ\Delta(\rho)\neq\Delta\circ\Lambda(\rho)$ for the state $\rho=\frac{1}{\sqrt{2}}(\ket{1}+\ket{2})$.

Applying the constructed operation $\Lambda(\cdot)$ in Eq. (\ref{incoherent_operation}) to the input state $\rho_{\mathrm{in}}$ defined in Eq. (\ref{input_state}) yields the output state
\begin{eqnarray}
\rho_{\mathrm{out}}=
\begin{pmatrix}
\frac34+\frac{7\sqrt2}{80} &
\frac14-\frac{i\sqrt2}{32} &
0 \\[4pt]
\frac14+\frac{i\sqrt2}{32} &
\frac14-\frac{7\sqrt2}{80} &
0 \\[4pt]
0 & 0 & 0
\end{pmatrix}.
\label{output_state}
\end{eqnarray}
A direct numerical evaluation of the coherence measure $C_m(\rho)$ for $\rho_{\mathrm{out}}$ yields
\begin{eqnarray}
C_m(\rho_{\mathrm{out}}) \approx 0.2548.
\end{eqnarray}
These values clearly demonstrate that $C_m(\rho)$ can increase under the action of the IO in Eq. (\ref{incoherent_operation}), namely
\begin{eqnarray}
C_m(\rho_{\mathrm{out}})-C_m(\rho_{\mathrm{in}})\approx +0.0695
\end{eqnarray}
Because the monotonicity of $C_m(\rho)$ under SIOs and DIOs forbids any increase in this measure, an IO-induced growth of $C_m(\rho)$ directly implies that the corresponding state transformation cannot be achieved within SIOs and DIOs. Hence, the explicit increase from $C_m(\rho_{\text{in}})$ to $C_m(\rho_{\text{out}})$ under the constructed IO demonstrates that the transformation from $\rho_{\text{in}}$ to $\rho_{\text{out}}$ can be achieved solely by IOs and remains impossible for any SIO or DIO. This completes the proof of the theorem. \hfill $\blacksquare$

From Theorem 1, the following key implications can be drawn:

(i) The explicit construction presented in Theorem 1 resolves the long-standing question posed by Chitambar and Gour in Refs. \cite{Chitambar2016PRA,Chitambar2016PRL} by demonstrating that certain quantum state transformations are achievable within IOs but impossible under both SIOs and DIOs. An immediate consequence of this construction is that \(C_m(\rho)\) cannot serve as a coherence monotone under IOs. Since \(\mathrm{IO}\subseteq\mathrm{MIO}\), this failure of monotonicity extends to MIOs as well, despite the fact that \(C_m(\rho)\) is known to be an MIO monotone for qubit systems \cite{Chitambar2016PRL}. In particular, the explicit IO in Eq. (\ref{incoherent_operation}) converts \(\rho_1\) into \(\rho_2\) while increasing the value of \(C_m(\rho)\), i.e., \(C_m(\rho_2)>C_m(\rho_1)\), thereby providing a concrete violation of MIO monotonicity. These observations clarify the limitations of \(C_m(\rho)\) as a coherence quantifier beyond the qubit regime. For completeness, Table \ref{table} summarizes the monotonicity behavior of \(C_m(\rho)\) alongside several standard coherence measures under different classes of incoherent operations.

\begin{table*}
\begin{tabular}{c|c|c|c|c}
    \hline
    \hline
             &MIOs &  IOs    &  DIOs  & SIOs  \\
    \hline
    $C_{l_1}(\rho)$ & $\times$ ~~Ref. \cite{Bu2017QIC}& $\checkmark$ ~~Ref. \cite{Baumgratz2014PRL} & $\times$ ~~Ref. \cite{Bu2017QIC}   & $\checkmark$~~Ref.~\cite{Yadin2016PRX} \\
    \hline
    $C_r(\rho)$ &$\checkmark$ ~~Ref.~\cite{Aberg2006AX}& $\checkmark$ ~~Ref. \cite{Baumgratz2014PRL}  &  $\checkmark$ ~~Ref.~\cite{Marvian2016PRA}& $\checkmark$~~Ref.~\cite{Yadin2016PRX} \\
    \hline
    $C_R(\rho)$ &$\checkmark$ ~~Ref.~\cite{Napoli2016PRL}&  $\checkmark$~~Ref.~\cite{Napoli2016PRL}  &  $\checkmark$~~Ref.~\cite{Napoli2016PRL}& $\checkmark$~~Ref.~\cite{Napoli2016PRL}  \\
     \hline
    $C_m(\rho)$& $\times$~~Theorem 1 &  $\times$~~Theorem 1   &  $\checkmark$  ~~Ref.~\cite{Chitambar2016PRA}   & $\checkmark$ ~~Ref.~\cite{Liu2023PRA} \\
    \hline
    \hline
    \end{tabular}
\caption{
The table shows the monotonicity of four coherence measures under three classes of operations: maximally incoherent operations (MIOs), incoherent operations (IOs), dephasing-covariant incoherent operations (DIOs), and strictly incoherent operations (SIOs). The coherence measures are the \( \ell_1 \) norm \(C_{I_1}(\rho)\), the relative entropy \(C_r(\rho)\), the robustness \(C_R(\rho)\), and the measure \(C_m(\rho)\) from Eq. (\ref{maximum}), where a checkmark (\(\checkmark\)) indicates monotonicity and a cross (\(\times\)) its absence. }\label{table}
\end{table*}

(ii) The quantity \(C_m(\rho)\) characterizes the maximal fidelity with which a state \(\rho\) can be converted, under stochastic SIOs, into the $d$-dimensional maximally coherent state \(\ket{\psi_d}=\frac{1}{\sqrt d}\sum_{j=0}^{d-1}\ket{j}\) \cite{Liu2023PRA}. A stochastic SIO is a completely positive, trace-nonincreasing map \(\Lambda_s(\cdot)=\sum_\mu K_\mu(\cdot)K_\mu^\dagger\) whose Kraus operators \(K_\mu\) are strictly incoherent. Theorem 1 shows that this SIO-based bound can be surpassed once IOs are allowed. For instance, applying the stochastic IO \(\Lambda_s(\rho)=K_2\rho K_2^\dagger/\mathrm{Tr}(K_2\rho K_2^\dagger)\) yields
\begin{eqnarray}
\rho_{\mathrm{out}}=
\begin{pmatrix}
\frac{40+4\sqrt{2}}{60-\sqrt{2}} & \frac{20-3i}{60-\sqrt{2}} & 0 \\[4pt]
\frac{20+3i}{60-\sqrt{2}} & \frac{20-5\sqrt2}{60-\sqrt{2}} & 0 \\[4pt]
0 & 0 & 0
\end{pmatrix},
\end{eqnarray}
for which \(C_m(\rho_{\mathrm{out}})\approx 0.3153\). This explicitly demonstrates that stochastic IOs can achieve values of \(C_m(\rho)\) exceeding the maximal limit established in Ref. \cite{Liu2023PRA}.

\section{Coherence monotones and state transformations}
Motivated by Theorem 1, which reveals a sharp separation between different classes of free operations, we now examine how this distinction manifests at the level of coherence monotones and state convertibility. In any resource theory, these two components, the monotones that quantify the resource and the operational criteria that determine when one state can be transformed into another, are intimately linked. This connection has been emphasized in recent works \cite{Datta2023PRL, Scandi2021PRL}, which ask whether one can identify a complete family of coherence monotones that is both necessary and sufficient to characterize convertibility under a given class of free operations. Formally, a set of monotones \({M_j(\rho)}\) is said to be complete if the transformation \(\rho_1 \to \rho_2\) is achievable under the free operations exactly when \(M_j(\rho_1) \ge M_j(\rho_2)\) for all \(j\).

For qubit systems, the situation is notably simpler. It is known that MIOs, DIOs, IOs, and SIOs possess identical transformational power for single-qubit states, and that qubit convertibility can be fully characterized by two monotones \cite{Chitambar2016PRA,Chitambar2016PRL}: the robustness of coherence \cite{Napoli2016PRL} \begin{eqnarray}
C_R(\rho)=\min_{\tau}\left\{s \ge 0|\frac{\rho + s \tau}{1 + s} \in \mathcal{I}\right\}
\end{eqnarray}
and \(C_m(\rho)\) in Eq. (\ref{maximum}) \cite{note}. Specifically, a qubit transformation from \(\rho_1\) to \(\rho_2\) is achievable under any of these operation classes if and only if
\begin{eqnarray}
C_R(\rho_1)\ge C_R(\rho_2)~~\text{and}~~C_m(\rho_1)\ge C_m(\rho_2).
\end{eqnarray}
This complete characterization for the qubit case motivates the deeper investigation carried out here.

Next, we shall show that when considering problems in more than two dimensions, the above issue becomes more complex.

First, by considering the complete set of monotones under SIO, we analyze the necessity of the complement \(\mathcal{M}_{\mathrm{SIO}} \setminus \mathcal{M}_{\mathrm{IO}}\) (i.e., the set of SIO monotones that do not belong to IO). Here and in what follows, \(\mathcal{M}_{\mathcal{F}}\) denotes the set of coherence monotones under a given class \(\mathcal{F}\) of free operations.

We begin by showing that $\mathcal{M}_{\mathrm{IO}} \subset \mathcal{M}_{\mathrm{SIO}}$. To see this, suppose \(M(\rho)\) is a coherence monotone under IOs, i.e., there is \(M(\rho) \ge M(\Lambda_{\mathrm{IO}}(\rho))\). Since \(\mathrm{SIO}\subset \mathrm{IO}\), it follows immediately that \(M(\rho)\) must also be a coherence monotone under SIOs \(M(\rho) \ge M(\Lambda_{\mathrm{SIO}}(\rho))\). Due to this reason, most of the commonly used coherence measures under SIOs are inherited from IO measures, such as the $l_1$ norm of coherence $C_{l_1}(\rho)$ \cite{Baumgratz2014PRL}, the relative entropy of coherence \cite{Baumgratz2014PRL}, and the robustness of coherence \cite{Napoli2016PRL}.

However, we shall next show that even when all IO monotones are considered, they still do not form a complete set under SIOs. In other words, there exist two states $\rho_1$ and $\rho_2$ for which, even though $M(\rho_1) \geq M(\rho_2)$ holds for every IO monotone $M(\rho)$, it does not imply that the transformation from $\rho_1$ to $\rho_2$ can be achieved via SIOs. This leads to the following Theorem 2.

\emph{Theorem 2}.--There exist two quantum states \(\rho_1\) and \(\rho_2\) such that \(M(\rho_1) \geq M(\rho_2)\) for every IO monotone \(M(\rho)\), although the transformation from \(\rho_1\) to \(\rho_2\) cannot be realized by any SIO.

\emph{Proof}.--Let \(\rho_1\) and \(\rho_2\) be the states defined in Eqs. (\ref{input_state}) and (\ref{output_state}), respectively. As shown by the explicit incoherent operation given in Eq. (\ref{incoherent_operation}), the transformation \(\rho_1 \rightarrow \rho_2\) is achievable within the class of IOs. Since any IO monotone \(M\) is non-increasing under IOs, it follows that
\begin{eqnarray}
M(\rho_1) \geq M(\rho_2)
\end{eqnarray}
for all coherence monotones under IOs.

On the other hand, according to Theorem 1, the transformation from \(\rho_1\) to \(\rho_2\) is impossible under SIOs. Therefore, while all IO coherence monotones agree that \(\rho_1\) is at least as coherent as \(\rho_2\), the transformation is nevertheless forbidden within the strictly smaller class of SIOs. This completes the proof of the theorem. \hfill $\blacksquare$

Theorem 2 thus indicates that a complete set of monotones under SIO must include coherence monotones belonging exclusively to \(\mathcal{M}_{\mathrm{SIO}} \setminus \mathcal{M}_{\mathrm{IO}}\).

In passing, it is important to recognize that previous analyses in Refs. \cite{Datta2023PRL,Du2019PRA} have shown any finite set of coherence monotones to be insufficient for completeness under SIOs. In contrast, the class \(\mathcal{M}_{\mathrm{IO}}\) considered in Theorem 2 contains infinitely many such monotones. In fact, \(\mathcal{M}_{\mathrm{IO}}\) encompasses all convex roof coherence measures \cite{Du2015QIC,Zhu2017PRA}. Although convex roof coherence measures provide a complete characterization of pure-state transformations under SIOs \cite{Zhu2017PRA}, Theorem 2 shows that this characterization does not extend to mixed-state transformations.

\begin{figure}[h]
\includegraphics[width=7cm]{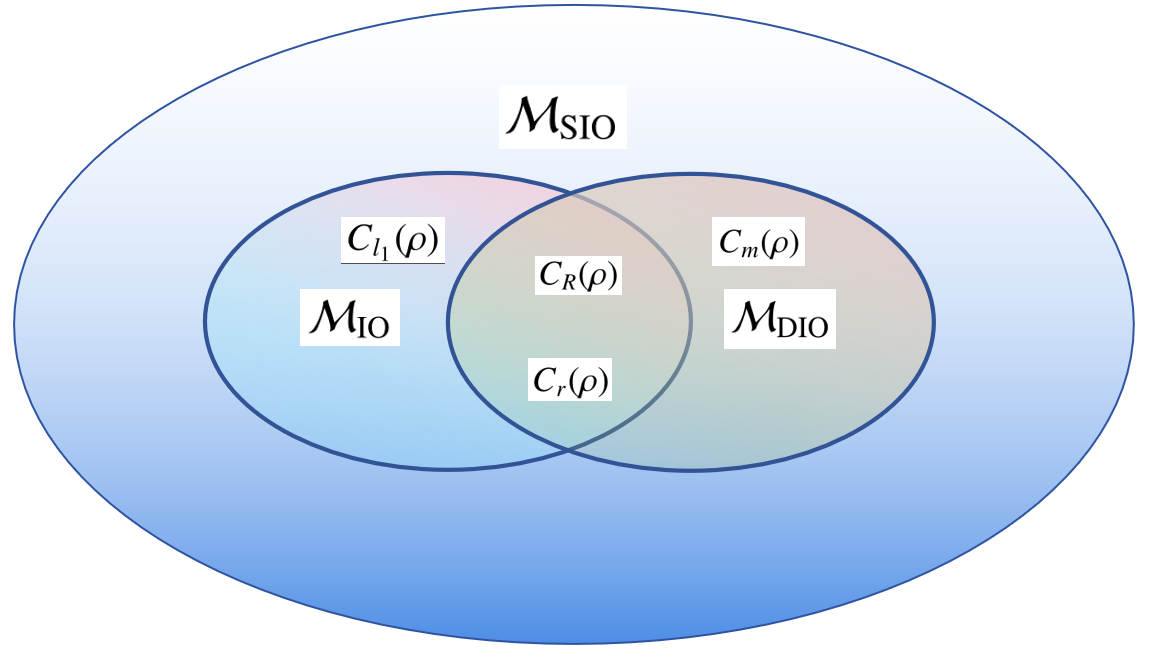}
\centering
\caption{Schematic of the inclusion relations among the sets of coherence monotones under different classes of free operations. Here, \(\mathcal{M}_{\mathcal{F}}\) denotes the set of coherence monotones under a given class \(\mathcal{F}\) of free operations. }
\label{fig1}
\end{figure}

A similar discussion applies to the analysis of DIOs, provided \(\mathcal{M}_{\mathrm{IO}}\) is replaced by \(\mathcal{M}_{\mathrm{IO}} \bigcap \mathcal{M}_{\mathrm{DIO}}\).
These set-theoretic relationships are illustrated schematically in Fig. \ref{fig1}. The figure shows that although \(\mathcal{M}_{\text{IO}}\) and \(\mathcal{M}_{\text{DIO}}\) are both subsets of \(\mathcal{M}_{\text{SIO}}\), neither class contains the other. Each class possesses coherence monotones that are distinct to that class, as summarized in Table \ref{table}. For example, the \(l_1\)-norm of coherence \(C_{l_1}(\rho)\) belongs to \(\mathcal{M}_{\text{IO}}\) but not to \(\mathcal{M}_{\text{DIO}}\), whereas the monotone \(C_{m}(\rho)\) is contained in \(\mathcal{M}_{\text{DIO}}\) but not in \(\mathcal{M}_{\text{IO}}\).

Based on the notions introduced above, we now state the following result, which emphasizes that one cannot restrict attention solely to those DIO monotones inherited from $\mathcal{M}_{\mathrm{IO}}$. The proof proceeds analogously to that of Theorem 2 and is therefore omitted. This leads to Theorem 3.

\emph{Theorem 3}.--There exist two quantum states \(\rho_1\) and \(\rho_2\) such that \(M(\rho_1) \geq M(\rho_2)\) for every coherence monotone $M(\rho)$ in \(\mathcal{M}_{\mathrm{IO}} \bigcap \mathcal{M}_{\mathrm{DIO}}\), although the transformation from \(\rho_1\) to \(\rho_2\) cannot be realized by any DIO.

\section{Conclusions}
In summary, we have explicitly constructed quantum state transformations that are achievable under incoherent operations but impossible under both strictly incoherent operations and dephasing-covariant incoherent operations, as shown in Theorem 1. This provides a definitive answer to whether incoherent operations can outperform dephasing-covariant operations, establishing a refined operational hierarchy among these classes. Our results further highlight fundamental limitations of monotone-based frameworks. Even a complete set of coherence monotones under incoherent operations cannot fully characterize state convertibility under the more restrictive strictly incoherent operations, as demonstrated in Theorem 2. More subtly, monotones that are valid under both incoherent and dephasing-covariant operations remain insufficient to determine convertibility under dephasing-covariant operations alone, as shown in Theorem 3. These observations underscore that operational criteria are indispensable for assessing transformations under specific free operations, and that a complete resource theory for any class must incorporate monotones that may not be valid under broader classes. Overall, our findings advance the understanding of quantum coherence manipulation by clarifying the intricate relationship between operational power and the associated sets of monotones for different classes of free operations.

\begin{acknowledgments}
This work was supported by the National Natural Science Foundation of China (Grant No. 12405019) and the Natural Science Foundation of Shandong Province (Grant No. ZR2025LLZ002).
\end{acknowledgments}


\begin{thebibliography}{99}
\bibitem{note} Although the expression \(C_{\Delta,R}(\rho)=\uplambda_{\max}\!\Bigl(\Delta\rho^{-1/2}\rho\,\Delta\rho^{-1/2}\Bigr) \) is used in Ref. \cite{Chitambar2016PRL}, since it is a monotonic function of \( C_m(\rho) \), we do not distinguish between them here.
\end{thebibliography}
\end{document}